\documentclass[12pt,epsf]{article}

\usepackage{amsmath}
\usepackage{amssymb}
\usepackage{makeidx}
\usepackage{epsf}
\usepackage{graphicx}        % standard LaTeX graphics tool
\makeindex

\def\inbar{\vrule height1.5ex width.4pt depth0pt}
\def\IB{\relax{\rm I\kern-.18em B}}
\def\IC{\relax\,\hbox{$\inbar\kern-.3em{\rm C}$}}
\def\ID{\relax{\rm I\kern-.18em D}}
\def\IE{\relax{\rm I\kern-.18em E}}
\def\IF{\relax{\rm I\kern-.18em F}}
\def\IG{\relax\,\hbox{$\inbar\kern-.3em{\rm G}$}}
\def\IH{\relax{\rm I\kern-.18em H}}
\def\II{\relax{\rm I\kern-.18em I}}
\def\IK{\relax{\rm I\kern-.18em K}}
\def\IL{\relax{\rm I\kern-.18em L}}
\def\IM{\relax{\rm I\kern-.18em M}}
\def\IN{\relax{\rm I\kern-.18em N}}
\def\IO{\relax\,\hbox{$\inbar\kern-.3em{\rm O}$}}
\def\IP{\relax{\rm I\kern-.18em P}}
\def\IQ{\relax\,\hbox{$\inbar\kern-.3em{\rm Q}$}}
\def\IR{\relax{\rm I\kern-.18em R}}
\def\IT{\relax{\rm I\kern-.18em T}}
\def\ZZ{\relax{\sf Z\kern-.4em Z}}

\def\a{\alpha}   \def\b{\beta}    \def\g{\gamma}  
 \def\G{\Gamma}     
    \def\Om{\Omega} \def\si{\sigma}
  \def\t{\tau}

\def\cA{{\cal A}}

  \def\cO{{\cal O}}
\def\cP{{\cal P}}

      \def\t-hat{{\hat t}}

\def\rmA{{\rm A}} \def\rmB{{\rm B}}  
    \def\rmI{{\rm I}}

       \def\rmFr{{\rm Fr}} 
\def\rmGal{{\rm Gal}}            
       \def\rmII{{\rm II}} \def\rmIII{{\rm III}} 
 \def\rmIm{{\rm Im}}

           \def\rmSL{{\rm SL}}

           \def\rmcs{{\rm cs}}
\def\rmdeg{{\rm deg}}       \def\rmdim{{\rm dim}}

       \def\rmmod{{\rm mod}} \def\rmmon{{\rm mon}}

      \def\rmth{{\rm th}}
  \def\rmtr{{\rm tr}}

\def\rmFr{{\rm Fr}}   

  \def\mathF{{\mathbb F}}   \def\mathN{{\mathbb N}}
\def\mathP{{\mathbb P}}
\def\mathQ{{\mathbb Q}}
\def\mathR{{\mathbb R}}     
\def\mathZ{{\mathbb Z}}

\def\fnote#1#2{\begingroup\def\thefootnote{#1}\footnote{#2}\addtocounter
{footnote}{-1}\endgroup}
\def\beq{\begin{equation}}
\def\eeq{\end{equation}}
\def\bea{\begin{eqnarray}}
\def\eea{\end{eqnarray}}

\def\lleq#1{\label{#1}\eeq}

\def\tabroom{\hbox to0pt{\phantom{\Huge A}\hss}}
\def\notin{\ \hbox{{$\in$}\kern-.51em\hbox{/}}}

\def\del{\partial} \def\vphi{\varphi}

  \def\E1Fq{E_1/\IF_q}

\def\notdiv{{\relax{~|\kern-.34em /~}}}

\def\del{{\partial}}

         \def\otau{{\overline \tau}}

\def\boxit#1{
\vbox{\hrule height1pt\hbox{\vrule width1pt\kern0.3cm
\vbox{\kern0.3cm\hbox{$\displaystyle#1$}\kern0.3cm}\kern0.3cm\vrule
width1pt}\hrule height1pt}}

\begin{document}

\phantom{ \hfill  \hfill \today~}

\vskip .9truein

\parindent=0pt
\baselineskip=16pt

\centerline{\large {\bf On Flux Vacua and Modularity}} 

\vskip .4truein

\centerline {\sc Rolf Schimmrigk\fnote{1}{rschimmr@iusb.edu; netahu@yahoo.com}}

\vskip .2truein

\centerline{Dept. of Physics}

\centerline{Indiana University South Bend}

\centerline{1700 Mishawaka Ave, South Bend, IN 46634}

\vskip 1.2truein

\baselineskip=18pt

\centerline{\bf Abstract}
\begin{quote}

Geometric modularity has recently been conjectured to be a characteristic feature for flux vacua with $W=0$. 
This paper provides support for the conjecture by computing motivic modular forms in a direct way 
for several string compactifications for which such vacua are  known to exist. The analysis of some 
Calabi-Yau manifolds which do not admit supersymmetric flux vacua shows that the reverse of the 
conjecture does not hold.
\end{quote}

\renewcommand\thepage{}
\newpage
\parindent=0pt

 \pagenumbering{arabic}

\tableofcontents

\vskip .5truein

\baselineskip=19.5pt
\parskip=.1truein

\section{Introduction}

Modularity is a theme that arises in quite different ways in string theory. From the early days of its development 
modular invariance on the worldsheet has been one of the cornerstones of the foundations of the theory. 
Historically far removed and much older than twodimensional 
conformal field theory is the notion of geometric modularity, which can be traced to Klein in the late 19th century, 
but which emerged as a more central part in mathematics only in the 1950s and 60s. First steps in this 
direction were taken in the work of Eichler, Taniyama, Shimura and Weil  \cite{w67} on elliptic curves, 
which eventually led to the insights of Langlands \cite{l80, l79} concerning higher dimensions. 
The latter work was originally aimed at 
a nonabelian generalization of class field theory, i.e. a nonabelian generalization of the relation between Artin's 
Galois theoretic $L$-functions and Hecke's modular $L$-series, but today the Langlands program has absorbed 
Grothendieck's notion of motives and subsumes in particular the grand conjecture that all motives are automorphic.
In a first approximation, motives can simply be thought of as subsectors of the cohomology of the variety.

The definition of automorphic forms is not canonical in the mathematics literature and different objects  are called 
automorphic. One clear distinction that can be drawn is between modularity in the classical sense of Klein and Hecke, 
which views modular forms as objects that are associated to the group $\rmSL(2,\mathR)$, and automorphic forms, which 
are associated to higher rank groups. It is this latter class of objects that the Langlands conjectures are really concerned 
with, and in the 
present paper the distinction between modular and automorphic forms will be made along these lines. 

The notion of geometric modularity did not play a role in the first exploration of string theory compactifications in the 
1980s and 1990s, perhaps because Langlands' conjectures are not very precise, they are computationally 
not immediately accessible, and more importantly, a physical interpretation of the purported modular and automorphic 
forms was lacking.  A first idea for such a physical interpretation came from the question
 whether geometrically induced modular forms can be related to  string theoretic 
modular forms on the worldsheet. This was pursued in a march through the dimensions, 
starting with the simplest possible string compactifications of complex dimension one \cite{su02}. 
Extensions to higher dimensions were then constructed 
 for K3 surfaces \cite{rs06},  Calabi-Yau threefolds and fourfolds \cite{rs08, rs13}, as well as for Fano-type 
mirrors of rigid CYs \cite{kls08}. Modularity in families 
of CY varieties was explored in \cite{kls10}. Related work in this direction was done later in the context of elliptic 
compactifications in \cite{kw18, kw19}.

Recently  it was suggested that modularity might also serve as an indicator for the existence 
of supersymmetric flux vacua in the framework of Calabi-Yau varieties \cite{k20etal}. For flux compactifications there are 
cohomological constraints for the field $G_3=F_3-\tau H_3$, which are conjectured 
to lead to modular forms in the classical sense for  flux compactifications
 with vanishing superpotential $W$. 
 This was supported in \cite{k20etal} with computations of several points in the complex moduli space of the 
 two-parameter octic embedded in the 
 configuration $\mathP_{(1,1,2,2,2)}[8]$.  There are other prominent Calabi-Yau configurations that
  are known to admit such supersymmetric flux vacua 
 and it is of interest to test the conjecture beyond this octic.
 
 A second issue is whether the modularity conjecture can be reversed
  in the way suggested in ref.  \cite{k20etal}, where the authors note that 
  one can imagine running the conjecture  in reverse to use modularity results to find new supersymmetric 
  flux compactifications.
 The idea that modularity implies the existence of supersymmetric flux vacua would be very useful because modularity 
  is expected to be a common occurrence. It is however not universal, as the example of the quintic threefold already 
  shows, and in general the Langlands conjectures only suggest the existence of automorphic forms. Hence 
 modularity is selective in the sense that not every manifold leads to classical modular forms and it is of interest to 
 analyze manifolds that have been shown not to admit flux vacua with $W=0$.

The outline of this paper is as follows. Section 2 contains a description of the methods used to compute the 
motivic $L$-series derived from weighted projective hypersurfaces. Section 3 extends the analysis of \cite{k20etal}
to several prominent Calabi-Yau threefolds that have been considered in the literature and
 for which flux vacua are known to exist, adding also a remark about modular black hole attractors. 
Section 4 addresses the issue of the reverse of the flux vacua 
modularity conjecture, and Section 5 contains the conclusions.

\vskip .2truein
  
\section{Motivic $L$-series in weighted hypersurfaces}

This section describes a method that allows to efficiently identify rank two motives $M$ for manifolds with high dimensional 
cohomology groups and to compute their $L$-series $L(M,s)$. These $L$-series are used to check for modularity and 
read off the level of their modular groups. This method will then be applied in the remainder of the paper to several 
manifolds that have been of interest in the context of flux vacua.

There exist different methods that can be used to compute motivic $L$-series of weighted hypersurfaces. Most common 
in the mathematical literature is the $p$-adic approach, but in the following the emphasis will be on motivic methods developed 
in a series of papers that were aimed at relating the resulting geometrically induced 
modular forms to forms on the worldsheet. This was initiated in \cite{su02} in the  context of elliptic 
curves and extended to higher dimensions in \cite{rs06, rs08, kls08, rs13} and to families in \cite{kls10}. The advantage 
of this method is its directness and simplicity, allowing the computation of motivic $L$-series without computing 
the full cohomology with subsequent factorization.  In the case of elliptic curves the motivic framework is not necessary 
because of the simple structure of their cohomology, but it becomes important when considering 
K3 surfaces and higher dimensional varieties, as discussed in the above references. 
  In order to prepare for the physical issues discussed in this 
paper  it is useful to briefly recall some of the structures that 
enter the arithmetic $L$-series.

The conceptual starting point is the zeta function $Z(X/\mathF_p,t)$ of Artin, Schmidt, and Weil, defined as 
a series expansion that collects the cardinalities $N_{r,p}(X)$ of the variety $X$ defined over
 finite field extensions $\mathF_{p^r}$ of $\mathF_p$
as
\beq
 Z(X/\mathF_p,t) ~=~ \exp\left( \sum_{r\geq 1} \frac{N_{r,p}(X)}{r} t^r\right).
\lleq{zeta}
Dwork's unscheduled proof \cite{d60} of the rationality of the zeta function  
\beq
Z(X/\mathF_p,t) = Q_p(X,t)/R_p(X,t)
\eeq
in terms of polynomials $Q_p, R_p$ gives this series (\ref{zeta}) a finite form that makes it useful because it shows that a finite 
amount of computation determines the complete structure of $Z(X/\mathF_p,t)$. The detailed structure of $Q_p,R_p$ was
outlined by Weil \cite{w49} and proven by Grothendieck \cite{g64} as given in terms of individual factors $\cP_p^j(X,t)$ that are
associated to the cohomology groups of the variety, with degrees given by the dimension of the $j^\rmth$ group
 \beq
 \rmdeg ~\cP_p^j(X,t) ~=~ \rmdim ~H^j(X).
\eeq
 The numerator collects the factors arising from the odd-dimensional groups, while the denominator runs through the even
  dimensional cohomology:
\beq
 Q_p(X,t) ~=~ \prod_{j=0}^{n-1} \cP_p^{2j+1}(X,t), ~~~~~
 R_p(X,t) ~=~ \prod_{j=0}^n \cP_p^{2j}(X,t).
 \eeq
 
 The full cohomology group of complex deformations is often a high-dimensional object, hence the polynomials 
 $\cP_p^j(X,t)$  are not very useful and neither are their completely factorized forms. 
 This motivates the search for smaller building blocks, first introduced by Grothendieck as motives.
 There are several ways in which motives can be described. The most familiar approach 
 perhaps is via  cohomological  realizations using the standard cohomology groups, but this is not immediately
 useful in an automorphic context. An alternative formulation via the concept of correspondences provides a more 
 geometric picture that makes contact with the mathematical goal to construct an appropriate category of these objects.
 In the present paper, following \cite{rs08}, motives are viewed as representations of the Galois group 
 because this approach provides the 
 simplest and most effective approach.
 A more detailed intuitive description of motives can be found in the appendix. 
 
 The Galois theoretic framework of motives is based on the 
 idea that associated to a manifold $X$ is a number field $K_X$ and that the Galois group $\rmGal(K_X/\mathQ)$ of $K_X$ acts 
 on the cohomology.  In the following all Galois groups are of the type $\rmGal(K/\mathQ)$ and the reference to $\mathQ$ will 
 be dropped for simplicity of notation. The action of this Galois group is in general reducible, leading to a decomposition of the full 
 group into irreducible sectors. These sectors can be viewed as realizations of the motives.
 A detailed introduction to general motives a la Grothendieck can be found in a string physics context in ref. \cite{rs08},
 which also contains references to the standard mathematical literature on motives. This paper also describes 
  the concrete realization of these objects that allows to specialize the abstract categorical 
treatment of the mathematics literature to concrete computations in the case of weighted hypersurfaces. 
 
  In the case of the manifolds of interest here the abstract definition of motives just outlined can be made concrete.
  The  important factor comes from the intermediate cohomology and can be written as
\beq
 \cP_p(X,t) ~=~ \prod_\a\left( 1+ j_{p}(\a) t\right) 
 \eeq
 where the $\a$ are rational vectors of dimension five
 that essentially parametrize the cohomology for certain primes $p$. More generally, the
  set of $\a$ for a weighted hypersurface of Brieskorn-Pham type of complex dimension $n$
   with weights $(w_0,...,w_{n+1})$ and degree $d$ is given by
\beq
 \cA_n^{d,p} :=
 \{\a \in \mathQ^{n+2}~{\Big |}~ 0 <\a_i <1, ~r_i=(d_i,p-1), r_i\a_i \equiv 0(\rmmod~1),
  \sum_i \a_i =0(\rmmod~1)\},
 \eeq
 where $d_i = d/w_i$.
 
 For a vector $\a$ the Jacobi sum is determined in terms of characters $\chi_{\a_i}(u)$ on finite fields 
 $\mathF_p$ which are  defined as
  \beq 
  \chi_{\a_i}(u)~ =~ e^{2\pi i m\a_i},
 \eeq
 where $m$ is determined in terms of the generator $g$ of $\mathF_p$ as $u=g^m$. With these characters
 the Jacobi sums can be written as
 \beq
  j_p(\a)
 ~=~ \frac{1}{p-1} \sum_{\stackrel{u_0+\cdots + u_{n+1}=0}{u_i \in \IF_p}}
     \chi_{\a_0}(u_0)\cdots \chi_{\a_{n+1}}(u_{n+1}).
 \eeq
 
 As noted above, the full cohomology group of complex deformations of weighted hypersurfaces is often a 
 high-dimensional space and it is more efficient to think in terms of the arithmetic building blocks of the manifolds.
 The simplification that arises from the motivic structure 
 is that, given any of the vectors $\a$ and the Galois group $\rmGal(K_X)$ of the number field $K_X$
  associated to the manifold $X$, we can consider 
 motives generated by the orbits $\cO_\a$ of $\a$ by the action of this Galois group on the vector $\a$.
 This leads to  a computationally useful representation of the motive as
 \beq
  M_\a ~=~ \rmGal(K_X) \cdot \a.
 \lleq{motive}
 
While for general varieties $X$ the 
field $K_X$ of the variety is determined by the factorization of the polynomials $\cP_p$ that enter the Dwork-Grothendieck 
factorization of the zeta function, for weighted hypersurfaces this field is immediately determined as the cyclotomic field 
$\mathQ(\mu_d)$, where $\mu_d$ is the cyclic group determined by the degree $d$ of the manifold.  The rank of the 
resulting motives is generically given by the order of the associated Galois group $\rmGal(\mathQ(\mu_d))$, which is isomorphic 
to $(\mathZ/d\mathZ)^\times$, and whose order is given by Euler's totient function $\phi(d)$, which can be computed via the 
product formula $\phi(d) = d\prod_{p|d} (p-1)/p$, where the product is over all prime divisors of the degree $d$. 
Hence the generic rank of the motive is 
determined by the degree of the hypersurface.
 
 The $L$-series of this motive $M_\a$ can then be obtained via the motivic polynomials 
 \beq
  \cP_p(M_\a, t) ~=~ \prod_{\b \in M_\a} \left(1+ j_{p} (\b) t \right) 
  \eeq
  as
  \beq
   L(M_\a, s) ~=~ \sum_n \frac{a_n(M_\a)}{n^s} ~=~ \prod_p \frac{1}{\cP_p(M_\a, p^{-s})}.
   \eeq
 With the motive $M_\a$ defined as the Galois orbit (\ref{motive}) the combination of Jacobi sums that determines 
 the coefficients in the $L$-series are rational integers even though the Jacobi sums themselves are complex. A proof,
 based on the fact that the action of an element $g\in \rmGal(\mathQ(\mu_d))$ on $\a$ induces an action on the Jacobi sum
 $j_p(\a)$,  can be  found in \cite{rs06}.
    
 A second type of $L$-series are those obtained from modular forms $f(q)$ defined relative to Hecke congruence subgroups 
 $\G_0(N)$ of the modular group $\rmSL(2,\mathZ)$ via a tensor type transformation behavior (see e.g. \cite{rs08} for 
 a physical discussion).  Associated to modular forms are $L$-series via the Mellin transform, 
 which via the expansion $f(q) = \sum_n a_nq^n$ leads to the same type of series $L(f,s) = \sum_n a_n n^{-s}$. 
 The question becomes whether  these $L$-series match for some weight and level $N$.
  
 The main interest in the earlier work cited above was to apply this construction to the holomorphic threeform $\Om$,
 which leads to the concept of the $\Om$-motive \cite{rs06} (see also \cite{rs08}), 
 and to relate the resulting modular and automorphic forms
 to the modular structure on the worldsheet. In the present paper the focus is on the existence of rank two motives 
 and their modularity. The focus in particular will be on Calabi-Yau varieties that have been of interest in the context 
 of flux compactifications.

An application of the arithmetic of CYs that is aimed at the connection with the string worldsheet conformal field theory, but 
is independent of the modular structure, can be found in ref. \cite{rs01}.
 
 \vskip .2truein
 
\section{Flux vacua with modular forms} 

Flux vacua with contributions from both the  RR field $F_{(3)}$ and the NSNS field $H_{(3)}$ in type IIB theory lead 
to a superpotential  $W$ that is usually written in terms of the complex axion-dilaton $\tau=C_0 + ie^{-\vphi}$ and the field 
 $G_{(3)} = F_{(3)} - \tau H_{(3)}$ as
  \beq
   W ~=~ \int_X G_{(3)}\wedge \Om,
  \eeq
  where $\Om$ is the holomorphic threeform. The specific form of $\tau$ adopted here is motivated by the fact that it lives
  on the upper halfplane, $\rmIm(\tau)>0$.

The vacuum constraints for the complex deformations $z^a$, defined as the $\Om$-periods via a homology basis $\{A_a, B^b\}$ 
and its dual,  as well as the axion-dilaton $\tau$, are given by 
\beq
 D_\tau W ~=~ 0 ~=~ D_aW,
 \lleq{critical}
 where $D_\tau W\equiv \del_\tau W + W \del_\tau K$ and $D_aW = \del_a W + W\del_a K$, where
  $K=K_\tau+ K_\rmcs$ is the K\"ahler potential of $\tau$ and the complex deformations respectively
  \beq
   K_\tau ~=~ - \ln (-i(\tau-\otau)), ~~~~~K_\rmcs ~=~ - \ln \left(i \int_X \Om\wedge {\overline \Om}\right).
  \eeq
  These constraints are sometimes called F-flatness.  If in addition to the criticality constraints (\ref{critical}) the 
  superpotential vanishes as well, $W=0$,  the resulting vacua are called supersymmetric. 
  While the criticality constraints determine $G_3$ to be of type 
  $H^{2,1}\oplus H^{0,3}$, the vanishing of the potential imposes the further constraint $G_3 \in H^{2,1}$ \cite{drs99, gkp02}. 
   In-depth reviews of flux compactifications can be found in 
  \cite{grana05, c07, p08}.

\subsection{Flux vacua for one-parameter weighted hypersurfaces}

The class of smooth Calabi-Yau hypersurfaces in weighted projective space consists of four spaces, all of which 
were considered first in the context of flux vacua in \cite{d04etal}. It was shown there that of these four only 
the degree six hypersurface $X_3^6 \in \mathP_{(1,1,1,1,2)}[6]$ leads to flux vacua with $W=0$, while the 
remaining three, given by the quintic $X_3^5\in \mathP_4[5]$, the octic $X_4^{8\rmA} \in \mathP_{(1,1,1,1,4)}[8]$ 
and the degree ten hypersurface $X_3^{10} \in \mathP_{(1,1,1,2,5)}[10]$, do not. Thus, while all these 
manifolds share the property that they have the simplest possible K\"ahler sector with $h^{1,1}=1$, they 
behave quite differently. The reason for this is to be found in the fact that the number fields $K_X$ associated to 
these manifolds have different degrees in that the smooth degree six surface leads to a quadratic extension of the 
rationals $\mathQ$, while the remaining spaces have fields of higher degree.

Thus,  if modular motives of rank two exist for any of these three manifolds (at the relevant $\psi$) then it is 
established that modularity is not sufficient for the existence of supersymmetric flux vacua. It is therefore of interest
 to check whether among these manifolds there exist spaces that are modular in the sense discussed here. 
 In the case of the quintic threefold $X_3^5$ it follows from the fact that the Galois group 
 $\rmGal(K_{X_3^5})= (\mathZ/5\mathZ)^\times=\{1,2,3,4\}$
 has order four, in combination with the structure of $H^{2,1}$ cohomology, that there are no rank two motives,
 consistent with the modularity conjecture. Similarly, for the degree ten hypersurface the Galois group 
 also has order four, $\rmGal(K_{X_3^{10}}) = (\mathZ/10\mathZ)^\times = \{1,3,7,9\}$, and it follows again 
 from the cohomological structure that all motives are of rank four. Thus in both of these cases the manifold
 is at best automorphic of higher rank.
 A similar analysis shows that the same holds for the configurations $X_3^{14}\in \mathP_{(1,2,2,2,7)}[14]$
 and $X_3^{15} \in \mathP_{(1,3,3,3,5)}[15]$, which have
  $(h^{1,1}, h^{2,1})=(2,122)$ and $(h^{1,1}, h^{2,1}) = (3, 75)$, respectively.
 
 These considerations already provide additional support of the modularity conjecture and the 
 discussion leaves among the one-parameter manifolds the degree six manifold, which does admit 
 supersymmetric flux vacua, and the smooth octic, which does not. The first of these two will be computed 
 presently, while the octic $X_3^{8\rmA}$ will be considered in the next section.

\subsection{The smooth degree six hypersurface  $X_3^6\in \mathP_{(1,1,1,1,2)}[6]$}

An example that appeared early in the discussions of flux vacua based on the class of weighted hypersurfaces
 constructed in  \cite{cls90, ks92, krsk92} 
is the degree six Brieskorn-Pham manifold in the configuration  $\mathP_{(1,1,1,1,2)}[6]$, defined by
\beq
 X_3^6 ~=~ \left\{\sum_{i=0}^3 z_i^6 ~+~ z_4^3 ~=~ 0\right\} ~\subset ~ \mathP_{(1,1,1,1,2)}.
 \eeq
 This is a smooth  hypersurface, hence it has only one K\"ahler form inherited from the ambient space, $h^{1,1}=1$,
 and the cohomology $H^{2,1}(X)$ is of monomial type,  $h^{2,1}=103$. 
 It was shown in ref. \cite{d04etal}  that supersymmetric flux vacua exist for this manifold at the Landau-Ginzburg 
 point,  i.e. the constraints $D_a W = 0 = D_\tau W$
 can be solved as well as $W=0$. Further discussions of flux vacua derived from this manifold can be found 
 in a number of papers, including \cite{d05, g05}. 
  
  In the context of a string theoretic interpretation of geometric 
   modularity this variety was discussed briefly in \cite{rs08}, where the modular 
  structure was established at the Landau-Ginzburg point described by the  Brieskorn-Pham geometry.
  In ref. \cite{rs08} the focus was on the $\Om$-motive, spanned by $H^{3,0}\oplus H^{0,3}$ of this manifold, 
  while in the context of flux vacua the cohomology 
  sector $H^{2,1}(X)\oplus H^{1,2}(X)$ is of interest.  By making the set $A_n^{d,p}$ defined in the previous section
   explicit for this hypersurface it becomes clear that 
  modulo permutations there are twelve basic $\a$-vectors that lead to six different motivic $L$-series. 
  These six $L$-functions arrange themselves into three pairs $L_{A\pm}$, $A=\rmI, \rmII, \rmIII$, of series that differ 
  only in the signs of some of their coefficients, indicating that they are character twists of each other. 
  These motivic $L$-series can be identified with modular form $L$-series as follows. 
   Parametrizing an independent set of  motives by the $\a$-vectors
   \beq
   \a_{A+} ~\in ~\left\{ \left(\frac{1}{3},\frac{1}{3},\frac{1}{2},\frac{1}{2}, \frac{1}{3}\right), 
       \left(\frac{1}{6},\frac{1}{3},\frac{1}{3},\frac{1}{2}, \frac{2}{3}\right),
       \left(\frac{1}{6},\frac{1}{6},\frac{1}{2},\frac{1}{2}, \frac{2}{3}\right) \right\}
  \eeq
  and  \beq
   \a_{A-} ~\in ~\left\{ \left(\frac{1}{3},\frac{1}{3},\frac{1}{3},\frac{1}{3}, \frac{2}{3}\right), 
       \left(\frac{1}{6},\frac{1}{2},\frac{1}{2},\frac{1}{2}, \frac{1}{3}\right),
       \left(\frac{1}{6},\frac{1}{6},\frac{1}{3},\frac{2}{3}, \frac{2}{3}\right) \right\}
  \eeq
  leads to the set of associated  $L$-series  $L(M_{A\pm},s)$. The coefficients of these $L$-series are not directly 
  fundamental though because each coefficient at a prime $p$ contains $p$ as a factor.  It is therefore natural 
  to shift the argument  $s$ of the $L$-function in order to eliminate these $p$-factors. This shift corresponds 
  to a twist in Hodge degree of the cohomology and has been important already in \cite{rs08}, as well as in the 
  context of mirrors of rigid Calabi-Yau varieties considered in \cite{kls08}. In the latter context this twist 
  can involve higher powers of the primes $p$. A selection of the $L$-function coefficients $a_p$
   that are obtained by dividing out these prime factors is shown in the table 1.
    \begin{center}
  \begin{tabular}{l | c c c c c c c }
  Prime $p$                &7            &13      &19            &31             &37  &43  &61 \\
  \hline
   $a_p(M_{\rmI\pm})$   &$\pm 1$   &$+5$        &$\pm 7$   &$\pm 4$   &$+11$  &$\mp 8$   &$-1$ \tabroom \\
   $a_p(M_{\rmII \pm})$  &$\pm 4$  &$+2$        &$\mp 8$   &$\pm 4$   &$- 10$  &$\mp 8$  &$+14$  \tabroom \\
   $a_p(M_{\rmIII\pm})$  &$\pm 5$  &$-7$   &$\mp 1$   &$\mp 4$   &$-1$   &$\pm 8$    &$-13$  \tabroom \\
  \hline
  \end{tabular}
  \end{center}
  \centerline{{\bf Table 1.} ~{\it Coefficients of the motivic $L$-series $L(M_{A\pm}, s+1)$ of $X_3^6$.}}

\vskip .2truein

  A perusal of various databases, starting with the list of Cremona of weight two forms \cite{c97}, as well as the more extensive
  list of forms by Meyer \cite{m05}, allows an identification of these $L$-series as arising from modular forms 
  $f_{A\pm}(q)=\sum_n a_nq^n, q=e^{2\pi i \tau}$, in the sense that their $L$-series agree with the $L$-series of the motives
  \beq
   L(M_{A\pm}, s+1) ~=~ L(f_{A\pm}, s),
  \eeq
  where the shift in $s$ implements the twist just discussed.
  Here the $f_{A+}$ are modular forms at levels $N=432, 144, 108$, and the 
  $f_{A-}$ are modular forms of weight two at levels $N=27, 36, 432$, respectively.  
  All these forms are relative to the Hecke congruence group $\G_0(N)$. The identification via the level 
  $N$ is sufficient for $N=27, 36$ and $N=108$, but is not unique when the space of rational 
  forms at a given level has more than one dimension.
  This was taken into account by Cremona \cite{c97} by introducing further alphabetical counting labels, a strategy that 
  was adopted in a slightly different way by the $L$-function and modular form data base (LMFDB) \cite{lmfdb}). 
   This is of relevance in the present discussion at the levels $N=144, 432$. In the former case the modular form 
   has the Cremona label 144A(A) and the LMFDB designation 144.2.a.a.  
  In the latter case the two   forms $f_{\rmI+}$ and $f_{\rmIII-}$ both belong to the same space,  but the table of 
  coefficients shows that their $L$-series are not the same.  In Cremona's list the designations are 432A and 432B, respectively,
  while in the LMFDB  the corresponding labels are {\tt 432.2.a.e} and {\tt 432.2.a.d}.. 
  The modular forms at level 27 and 144 have been considered previously in a string theoretic context 
  in  refs \cite{su02} and \cite{rs05}, where  they were shown to  admit a worldsheet interpretation for 
  elliptic compactifications. It was explained in \cite{rs05} that the motives $M_{A\pm}$ are of complex multiplication type,
  hence are modular with forms that admit complex multiplication, a symmetry that imposes a certain sparseness
  of the nonvanishing coefficients $a_p$.  These forms arise because the rank two motives $M_{A\pm}$ are
   of complex multiplication  type, determined by algebraic Hecke characters by the work of Deligne \cite{d79}. 
   Such characters have been shown to be modular in the work of Hecke \cite{hecke-werke}.
  All the motives and their forms constructed in this paper are of complex multiplication type.
  
  The modular forms identified above can be 
  shown to arise from the genus ten and genus four curves that are embedded in the hypersurface $X_3^6$
  and whose $L$-function decomposes into $L$-functions of elliptic curves $E_A$ that lead to 
  the modular forms $f_A$.  Including the $L$-function of the 
  $\Om$-motive thus leads to a completely modular structure of the third cohomology group of the form
  \beq 
  L(H^3(X_3^6),s) ~=~ L(f_\Om,s) \prod_{A\pm} L(f_{A\pm},s-1)^{a_{A\pm}}.
  \eeq
 Here $a_{A\pm} \in \mathN$ describe the multiplicities of the motives and the modular form $f_\Om$ of the $\Om$-motive
   is a cusp form of weight four and level  $N=108$ relative to the Hecke congruence subgroup $\G_0(N)$, 
   i.e. $f_\Om\in S_4(\G_0(108))$. 
   
 This shows that the cohomology of the degree six threefold hypersurfaces leads to a  total of seven different modular 
 forms $f_\Om, f_{A\pm}$.   
Combining the flux vacua analysis of \cite{g03etal} with the modularity analysis above thus provides further support
for the flux vacua modularity conjecture of ref. \cite{k20etal}.

\subsection{A modular rank two attractor on $X_3^6$}

A third arithmetic theme in string theory is provided by black hole attractors, initially emphasized in the number theoretic 
context by Moore \cite{m98}, and further developed in the context of complex multiplication in ref. \cite{lps03,lss03}. 
Recently, an extensive search 
for rank two attractors was reported in ref. \cite{c20etal}. It is in this context worth noting that the modularity of the 
$\Om$-motive $M_\Om$ of the degree six hypersurface $X_3^6$, established in \cite{rs08} and recalled briefly 
in the previous section,
 in terms of a weight four cusp form at level $N=108$, i.e. $f(M_\Om(X_3^6), q) \in S_4(\G_0(108)$, shows that the rank two 
attractor derived from the smooth degree six hypersurface is modular in the classical sense.

The form $f(M_\Om(X_3^6), q)$ of $X_3^6$
 has complex multiplications, which provides a second way to recognize 
complex multiplication as a characteristic of at least some black hole attractors. This is independent of the 
approach in \cite{lps03, lss03} via the Shioda-Katsura decomposition of the cohomology of curves embedded in weighted 
hypersurfaces.  The Deligne conjecture, discussed in the context of black hole attractors in 
\cite{lps03}, has been proven for motives with complex multiplication \cite{b86}, hence leads to a relation between the periods 
of the variety and the $L$-functions values of $M_\Om$.  Since the black hole potential
can be written in terms of the periods it immediately follows from the work of \cite{lps03} and \cite{rs08} 
that the  entropy can be expressed in terms of these 
$L$-function values. This was recently made explicit in a different example in \cite{c20etal} and is 
further discussed in \cite{y20}.

\subsection{Flux vacua for two-parameter hypersurfaces} 

\subsubsection{The octic $X_3^{8\rmB} \in \mathP_{(1,1,2,2,2)}[8]$}

The resolved Brieskorn-Pham octic hypersurface 
\beq
 X_3^{8\rmB} ~=~ \left\{ z_0^8 + z_1^8 + z_2^4+z_3^4+z_4^4 ~=~ 0 \right\} ~\subset ~ \mathP_{(1,1,2,2,2)}
 \eeq
 has Hodge numbers $h^{1,1}=2$ and $h^{2,1}=86$, where both sectors receive contributions of the resolution of the 
 singular set. This variety is a K3 fibration with generic fibers in the configuration of quartic surfaces $X_2^4 \in \mathP_3[4]$,
 which  has been considered in the context of 
 flux vacua in a number of papers, including \cite{d05,  b15etal, k20etal}. 
 In the latter reference this is the main configuration considered.
 
 Aspects of a string theoretic interpretation of geometrically induced modular forms of weight two  for this variety 
 were discussed in \cite{ls04}, where it was shown that this manifold leads to a weight two 
 modular form at level $N=64$. This modular form arises from the algebraic curve $C^4$ of degree four that defines the singular 
 set  embedded in this  manifold. While this curve has genus three, its $L$-function factorizes, in the process
  leading to an elliptic curve of degree four,  which is modular at the given level. 
  This form was also considered in the paper by Kachru, Nally and Yang \cite{k20etal} 
 as support for the modularity conjecture.
 
 In the motivic framework considered in this paper the starting point is the Galois group, which in the present 
 example has order four
 \beq
  \rmGal(K_{X_3^{8\rmB}}) ~\cong ~ \{1, 3,5,7\}.
 \eeq
 Hence the generic orbit has length four, which is in particular the case for the $\Om$-motive $M_\Om$. 
 These rank four motives are not modular but are expected to be automorphic according to the Langlands conjecture. 
 Having a Galois group 
 of order larger than two in general does not prevent the existence of orbits of shorter length, leading to the possibility of
 rank two motives that can be modular.  For the present case of the two-parameter octic 
 modular motives of rank two can be found, for example, at levels $N=32, 64$ for the motives parametrized
 by $\a=({\scriptsize \frac{1}{4}, \frac{1}{4}, \frac{1}{4}, \frac{1}{2}, \frac{3}{4} })$ and 
 $({\scriptsize \frac{1}{4}, \frac{1}{4}, \frac{1}{2}, \frac{1}{2}, \frac{1}{2} })$, 
 respectively, after implementing the Tate twist.
 
 The focus here has been on the Landau-Ginzburg point, geometrically given by the Brieskorn-Pham point in the 
 configuration. However, the observation just made, that higher order Galois groups can lead to rank two motives, 
 generalizes to the case of deformations of the variety away from the diagonal form. 
 This can be seen in the motivic framework by considering the theory of deformed motives in ref. \cite{kls10}.

 \subsubsection{The degree 12 hypersurface $X_3^{12}\in \mathP_{(1,1,2,2,6)}[12]$}
 
 A second two-parameter threefold that has been discussed extensively in the flux compactification context is 
  the resolved hypersurface 
 \beq
 X_3^{12} ~=~ \left\{ z_0^{12}+ z_1^{12} + z_2^6+z_3^6+z_4^2 ~=~ 0 \right\} ~\subset ~ \mathP_{(1,1,2,2,6)},
 \eeq
with $h^{1,1}=2$ and $h^{2,1}=128$. This is a K3 fibration with a typical smooth fiber in the configuration 
$X_2^6 \in \mathP_{(1,1,1,3)}[6]$ that  has been analyzed in refs. 
 \cite{g03etal, mn04, cq04, d05, c07etal}, and more recently in ref. \cite{c19etal}.  
 
 The order of the Galois group here is $\phi(12) = 4$, hence generically
 the motives will again be of rank four, much like in the case of the  octic hypersurface considered above. 
 Similar to the octic case there exist rank two motives for this variety as well.  It turns out that most of these
 rank two motives have already been encountered above in the discussion of the degree six threefold $X_3^6$
 even though the orders of the underlying Galois groups are different.
 As a result, their modularity follows immediately from the analysis of section 3.2. 
  The remaining rank two motives of this manifold can either be recovered from the 
  two-parameter model $X_3^{8\rmB}$
 considered above, or can be computed separately, leading to modular forms already encountered in
 our previous computations.
  
\vskip .2truein

\section{On the reverse of the modularity conjecture}

A natural question in the context of the flux vacua modularity conjecture is whether its reverse also holds, 
an issue that was addressed in  \cite{k20etal}. A positive answer to this question would be very interesting 
because it would imply that the existence of modular motives provides a diagnostic for the existence 
of flux vacua $D_\a W = 0 = D_\tau W$
  for which $W=0$. This can be discussed by testing the existence of modular rank two motives in 
  compactifications for which such $W=0$ do not exist.

As noted above already, it has been known for quite some time that  among the one-parameter models
 there are examples for which there exist no supersymmetric flux vacua, 
 i.e. there are points in the moduli space for which the flux dynamics 
  as well as $W=0$  cannot be solved \cite{d04etal}.  An example of this type that was analyzed in 
  \cite{g03etal, g04etal, d04etal, d05} is the smooth octic 
  hypersurface $X_3^{8A}\subset \mathP_{(1,1,1,1,4)}[8]$ at the Brieskorn-Pham point, or Landau-Ginzburg 
  point.  As also  noted above, it was furthermore shown 
  in \cite{d04etal, d05} that among the smooth weighted CY hypersurfaces
there is only one manifold that admits such supersymmetric flux vacua, namely the degree six hypersurface 
 $X_3^6 \in \mathP_{(1,1,1,1,2)}[6]$, analyzed in the previous section
while the remaining three do not. It was shown in the previous section that both the quintic and the degree ten 
hypersurface are not modular in the classical sense defined here, which leaves the smooth octic.

 \subsection{The one-parameter octic $X_3^{8\rmA}\in \mathP_{(1,1,1,1,4)}[8]$}
 
 The next space in the sequence of manifolds with $h^{1,1}=1$ is the smooth octic 
\beq
 X_3^{8A} = \left\{ p(z_i) ~=~ \sum_{i=0}^3 z_i^8 ~+~ z_4^2 ~=~ 0\right\} ~\subset ~ \mathP_{(1,1,1,1,4)}
\eeq
of Brieskorn-Pham type, with Hodge numbers $h^{1,1}=1$ and $h^{2,1}=149$. 
The orientifold projection is derived in \cite{g03etal} via Sen's method \cite{s97} 
from the CY fourfold $X_4^{24} \in \mathP_{(1,1,1,1,8,12)}[24]$, whose Euler number 
$\chi=23,328$ and cohomology has been computed in ref. \cite{lsw98}. The associated data
 can be found in  \cite{ellcy4list},  where it is the third model in the list of elliptic fibrations. 
In the context of flux vacua this manifold is also considered in \cite{g04etal, d04etal, d05}, and more 
recently in \cite{cs18, css19}.

The Galois group of the field $K_X$ of this degree eight hypersuface is of order four 
\beq
 \rmGal(K_X) ~=~ (\mathZ/8\mathZ)^\times ~=~ \{1,3,5,7\},
 \eeq
 hence the  $\Om$-motive of this variety is of rank four and is expected to be automorphic according to the Langlands 
conjectures. While the generic orbit in the intermediate cohomology thus has length four, in the context of flux vacua
 the interest is again in the existence of lower rank motives. In principle one can enumerate the motives as orbits, but the 
 main issue in the present discussion is whether there exist rank two modular motives in this variety.
  Examples of  two-dimensional  Galois orbits are generated by the vectors 
 \beq
 \a_+~=~ \left(\frac{1}{4},\frac{1}{4},\frac{1}{2},\frac{1}{2},\frac{1}{2}\right),~~~~~
 \a_- ~=~ \left(\frac{1}{4},\frac{1}{4},\frac{1}{4},\frac{3}{4},\frac{1}{2}\right) 
 \eeq
 hence  they parametrize rank two motives, denoted here by $M_\pm$. The computation of the motivic $L$-function 
 associated to $M_\pm$ along the  lines of section 2 leads to rescaled coefficients $a_p$ as collected in 
 table 2.
  \begin{center}
  \begin{tabular}{l | c c c c c c}
  Prime $p$   & 5  &13  &17  &29   &37   &41  \\
  \hline
  $a_p(M_\pm)$   &$\pm 2$  &$\mp 6$ &$+2$  &$\pm 10$  &$\pm 2$  &$+10$   \tabroom \\
  \hline
  \end{tabular}
  \end{center}
  \centerline{{\bf Table 2.} ~{\it $L$-series coefficients for the motives $M_\pm$ of $X_3^{8\rmA}$.}}
  
  \vskip .2truein
  
  The existing data bases again allow to identify these $L$-series as arising from weight two modular forms $f_\pm(q)$
 \beq
  L(f_\pm, s)  ~=~  1 ~\pm ~\frac{2}{5^s} - \frac{3}{9^s}  \mp \frac{6}{13^s} + \frac{2}{17^s} - \frac{1}{25^s}
      \pm  \frac{10}{29^s} ~\pm ~ \frac{2}{37^s} ~+~ \frac{10}{41^s} +   \cdots
  \lleq{motivic-L-series}
  which shows that the motives $M_\pm$ are modular after the Tate twist
  \beq
   L(M_\pm, s+1) ~=~ L(f_\pm, s).
  \eeq
  Here the  weight two modular forms $f_\pm$ are at levels $N=64$ and $N=32$, respectively. At these levels the 
 spaces of rational forms are one-dimensional, hence $N$ determines these forms uniquely. 
  These modular forms have appeared previously in the string 
theory analysis of \cite{ls04} and \cite{rs05}. 

The analysis here thus leads to the conclusion that the one-parameter octic Brieskorn-Pham hypersurface 
embedded in $\mathP_{(1,1,1,1,4)}$ 
is modular in the sense of carrying rank two motives that are modular in the classical sense, while not admitting 
supersymmetric flux vacua.

\subsection{The two-parameter hypersurface $X_3^{18}\in \mathP_{(1,1,1,6,9}[18]$}

The above analysis for $X_3^{8\rmA}$ can be applied to other weighted hypersurfaces that have been considered 
in the context of the existence/nonexistence of $W=0$ flux vacua. One example that was considered early on in many 
papers on flux vacua, and which has received recent attention,  is the degree 18 hypersurface 
\beq
 X_3^{18} ~=~ \left\{z_0^{18}+z_1^{18} + z_2^{18} + z_3^6 + z_4^2 ~=~ 0\right\} ~\subset \mathP_{(1,1,1,6,9)},
 \eeq
 see e.g. \cite{ddf04, b05etal, c05etal, d05, ds05,  b06etal, c07etal, a07etal, b09etal, l12etal, d19etal}.
 This is an elliptic fibration with $h^{1,1}=2$, $h^{2,1}=272$, hence $\chi = -540$, with a typical
  fiber in the configuration  $E^6 \in \mathP_{(1,2,3)}[6]$.
  An analysis of DeWolfe \cite{d05} based on R symmetries first showed that there are no $W=0$ flux vacua associated 
 to this manifold. It is therefore again of interest to consider the motivic structure of this manifold. In this case the 
 Galois group has order six 
  \beq
   \rmGal(K_{X_3^{18}}) ~=~ \{1, 5, 7, 11, 13, 17\}
  \eeq
  and hence the rank of the motives will  generically be six. This holds in particular for the $\Om$-motive, which therefore 
  is expected to be automorphic, not modular. However,  despite this larger Galois group this manifold does have rank two motives.
  
 As before, we consider the image of $\a$-vectors under the Galois group. An example of a rank two motive is given by 
  \beq
   M ~=~ \left(\frac{1}{3}, \frac{1}{3}, \frac{1}{2}, \frac{1}{3}, \frac{1}{2}\right) 
      ~\oplus ~\left(\frac{2}{3}, \frac{2}{3}, \frac{1}{2}, \frac{2}{3}, \frac{1}{2}\right).  
 \eeq
 The comparison of this motive with the rank two motive $M_{\rmI+}(X_3^6)$ of the degree six hypersurface
  $X_3^6 \in \mathP_{(1,1,1,1,2)}[6]$ considered above
 shows that $M$ is identical to $M_{\rmI+}$ even though the Galois groups have different orders. 
 It follows that the $L$-series  has the same expansion as in the case of the motive $M_{\rmI+}$ 
 of the degree six hypersurface, hence leads again to the level $N=432$ modular form designated by the LMFDB 
 as 432.2.a.e.  There are further rank two motives that arise in the manifold and these can be analyzed in a similar
 way.
 
\subsection{The three-parameter hypersurface $X_3^{24}\in \mathP_{(1,1,2,8,12)}[24]$}

As a final example consider the variety
\beq
 X_3^{24} ~=~ \left\{z_0^{24}+z_1^{24}+z_2^{12}+z_3^3 + z_4^2=0\right\}.
 \eeq
 This configuration is iteratively structured in that it is not only an elliptic fibration with typical fiber in the 
  configuration $E^6 \in \mathP_{(1,2,3)}[6]$,
 but also a K3-fibration with typical fiber in the configuration $X_2^{12}\in \mathP_{(1,1,4,6)}[12]$. 
 Its flux vacuum structure was considered in DeWolfe \cite{d05} as a three-parameter model with 
  Hodge numbers $h^{1,1}=3$ 
 and $h^{2,1}=243$, leading to a  K\"ahler structure that is more involved
 than the often considered two-parameter examples. Other work on this manifold 
 in the context of flux compactifications includes
 \cite{et05, pw14}. 
 
 The  manifold $X_3^{24}$ has the highest degree $d=24$ of the manifolds considered in this paper, 
 leading to the largest Galois group of all examples. Since the totient function here is 
 $\phi(24)=8$, a typical Galois orbit leads to a motive of rank eight. 
 %More precisely, the Galois group is isomorphic to 
 %\beq
  %\rmGal(X_3^{24}) ~=~ \{1,5,7,11,13,17,19,23\}.
  %\eeq
  It turns out that despite the large order of this group, 
 the modular structure of this manifold is in part similar to that of the degree six smooth hypersurface $X_3^6$
  because their motivic structure 
 partially overlaps. The modular analysis of section three 
 thus immediately leads to the existence of modular rank two motives for the manifold $X_3^{24}$.
 
\vskip .2truein
 
\section{Discussion}

In this paper the modular and automorphic structure of most of the Calabi-Yau manifolds considered 
in the context of flux compactifications has been established.  
The methods used here are direct and allow the computation of the necessary $L$-functions 
from the motive itself, rather than from the factorization of the full cohomology. 
These results have implications  for the modularity conjecture for supersymmetric flux vacua 
formulated in \cite{k20etal}.  Support for the conjecture 
has been provided above by establishing the existence or non-existence of  modular rank two motives for 
 all one-parameter smooth Calabi-Yau hypersurfaces of dimension three. 
 While the smooth degree six and eight weighted hypersurfaces admit such modular motives, the 
 quintic and degree ten hypersurfaces do not.
 
A second issue is raised by the question whether rank two modularity of motives
in the cohomology sector $H^{2,1}(X)\oplus H^{1,2}(X)$ can be used as an indicator for the existence 
of supersymmetric flux vacua. Since modularity is expected to be common this would lead to the expectation 
that the existence of supersymmetric flux vacua is in some sense also generic. This was first addressed here
by considering the one-parameter hypersurface of degree eight  in the weighted projective space $\mathP_{(1,1,1,1,4)}[8]$.
For this variety modular rank two motives can be constructed, thereby establishing that modularity is not a 
definitive diagnostic for the existence of flux vacua with vanishing superpotential.  Other manifolds that have been 
considered in the flux vacua literature have a similar structure. Among these are the degree 18 configuration 
$X_3^{18} \in \mathP_{(1,1,1,6,9)}[18]$ and the degree 24 configuration $X_3^{24} \in \mathP_{(1,1,2,8,12)}[24]$,
both of which were considered in the flux context in \cite{d05}, where it was shown that these two spaces do not 
admit flux vacua with vanishing superpotential. The first of these is a prominent two-parameter model,
while the latter has a three-dimensional K\"ahler sector. Both of these spaces have 
  rank two motives that are modular, similar to the one-parameter octic. 
 
 The results of this paper thus strengthen the modularity conjecture of ref. \cite{k20etal}  but establish that 
 modularity is not a necessary ingredient for the existence of supersymmetric flux vacua.

\vskip .3truein

\section{Appendix: Motives}

This appendix provides a brief summary of the conceptual background for 
the motivic constructions given in the main part of this paper. 

The theory of motives has its roots in the work of Hurwitz from 
the late 19th century and  Deuring from the 1930s, but as a conceptual framework it
was initiated by Grothendieck in the 1960s.  In the intervening decades
it has shown persistent resistance to a complete understanding. This is in particular the case
 for mixed motives that are important in some physical constructions \cite{kls10}, but do not play a role in the present paper. 
  For the discussion here it therefore  suffices  to focus on the theory of pure motives because the motives 
 that are considered in the main text are 
not affected by the resolution of the  weighted hypersurface.
Even in the case of pure motives the properties that are expected of these objects depend on conjectures formulated 
originally by Grothendieck that have not been proven.
An extensive discussion of the geometric theory of motives can be found in the book \cite{mnp13}, and some 
aspects of the arithmetic theory are discussed in ref.  \cite{a04}. More extensive reviews in a physical context 
can be found in \cite{rs08, k20etal}.  Most of the mathematical literature on 
motives is concerned with the categorical aspects of the theory.  This is important for the basic philosophy 
of what the theory of motives is supposed to achieve, and it is also  of interest for some physical applications 
\cite{rs06, rs08}. However, for the questions addressed in the present paper this 
is not necessary. The notation here is as light as possible in order to focus on the conceptual picture.

As mentioned in the main text, there are two main approaches to motives, one geometric, the other cohomological. 
In the geometric approach, based on the notion of a correspondence,  motives $M$ associated to a variety $X$ are 
defined as triplets $M=(X,\si, m)$, where $\si$ is a projector defined by an algebraic cycle 
$\si$ in $X\times X$ relative to an equivalence relation and $m$ is an  integer.  The concept of a projector makes sense
because correspondences can be combined and with an appropriately defined composition law they form a ring.
 There are several equivalence relations  that can be considered for algebraic cycles and relations between 
 different pure motives depend crucially  on which of these relations is considered, as well as on the integer $m$. 
 In category language this addresses the 
form of the morphisms between motives $(X, \si_X, m)$ and $(Y,\si_Y,n)$.  More important for this paper is that the 
integer $m$ determines the rescaling of the $L$-function encountered above via the Tate twist.

The arithmetic properties of such a pure motive $M$ are obtained by considering the intersection of $\si$ with the 
 graph $\G_{\rmFr_q^r}$ of the $r^\rmth$   iterate of the Frobenius map $\rmFr_q$ which sends a point $x$ to $x^q$,
  leading to 
  \beq
  N_{r,q}(M) = \langle ^t\si \cdot \G_{\rmFr_q^r} \rangle.
  \eeq

   By using the Lefschetz trace formula the cardinalities can alternatively be obtained by considering the action induced by
the Frobenius automorphism $\rmFr_q$ and its iterates on the ($\ell$-adic) cohomological realizations $H_\ell^i(M)$ of 
 the motive $M$, leading to the consideration of the traces $\rmtr(\rmFr_q^r{\Big |}H_\ell^i(M))$, where the induced 
 action is denoted by the same symbol as the  Frobenius map.  
 Here $H_\ell(M)$ is to be viewed as a Galois representation of some dimension that in general is different from the 
  dimension of the cohomology group $H^i(X)$ in any of its usual
  forms. This is, in brief, the outline of the intersection theoretic and cohomological approach as it applies 
   to the arithmetic theory of motives. 
In order to proceed one needs to fix a field $K$ so that the Galois group of $K$ can be used to construct motives. 
The last step is to obtain a description of the cohomology in such a form that the action of this Galois group is easy to deal with.

For a variety $X$ the Weil conjectures, proven by Grothendieck and Deligne, lead to a natural candidate of a 
field $K_X$ \cite{rs06, rs08}. First consider the Weil-Grothendieck decomposition of the zeta function 
$Z(X/\mathF_p,t)$ into its cohomological 
pieces $\cP_p^i(X,t)$ and then decompose these further $\cP_p^i(X,t) = \prod_j (1-\g_j^i(p) t^j)$.
The final Weil conjecture, proven by Deligne, states that the $\g_j^i$ are algebraic integers with a certain norm, 
hence they lead to an 
algebraic number field constructed as $K_X = \mathQ(\{\g_j^i\})$. In the case of the Brieskorn-Pham varieties considered 
above the $\g_j^i$ are the Jacobi sums and the number field simplifies in the way discussed earlier in section two.
The second simplification that arises for the varieties considered here is that if one considers  motives based on the monomial 
part $H^3_\rmmon(X)$ of the complex deformation cohomology $H^3(X)$ then an explicit decomposition  of this group 
can be constructed. This decomposition is parametrized by the $\a$-vectors considered above
and the action of the field $K_X=\mathQ(\mu_d)$ on the group $H_\rmmon^3(X)$ can be made explicit as an action on 
these vectors, as described in section two.

\vskip .3truein

{\large {\bf Acknowledgement.}} \\
It is a pleasure to thank  Ralph Blumenhagen, Ilka Brunner, and Monika Lynker for discussions and correspondence.
The appendix on an intuitive and brief description  of motives was added at the suggestion of the referee.

\vskip .4truein
%\vfill \eject


\baselineskip=17pt
\parskip=0pt
\begin{thebibliography}{}
 \bibitem{w67} A. Weil, {\it \"Uber die Bestimmung Dirichletscher Reihen durch Funktionalgleichungen}, 
     Math. Ann. {\bf168} (1967) 149 $-$ 156
 \bibitem{l80} R. Langlands, {\it L-Functions and automorphic representations}, Proc. Int. Cong. Math. 1978, 
 Helsinki, Acad. Sci. Fennica, 1980
 \bibitem{l79} R. Langlands, {\it Automorphic representations, Shimura varieties, and motives. Ein M\"archen}, 
   in Proc. of Symposia in Pure Mathematica {\bf 33} Part 2 (1979)  205 $-$ 246
 \bibitem{su02} R. Schimmrigk and S. Underwood, {\it The Shimura-Taniyama conjecture and conformal field theories}, 
    J. Geom. Phys. {\bf 48} (2003) 169 $-$ 189, arXiv: hep-th/0211284
\bibitem{rs06} R. Schimmrigk, {\it The Langlands program and string modular K3 surfaces}, 
     Nucl. Phys. {\bf B771} (2007) 143 $-$ 166, arXiv: hep-th/0603234 
 \bibitem{rs08}  R. Schimmrigk, {\it Emergent spacetime from modular motives}, 
  Commun. Math. Phys. {\bf 303} (2011) 1 $-$ 30, arXiv: 0812.4450 [hep-th]
 \bibitem{rs13} R. Schimmrigk, {\it String automorphic motives of nondiagonal varieties}, arXiv: 1301.2583 [hep-th]
 \bibitem{kls08} S. Kharel, M. Lynker and R. Schimmrigk, {\it String modular motives for mirrors of rigid Calabi-Yau varieties}, 
   Fields Inst. Commun. {\bf 54} (2008) 47 $-$ 63, arXiv: math.AG/0908.1256
  \bibitem{kls10} S. Kadir, M. Lynker and R. Schimmrigk, {\it String modular phases in Calabi-Yau families}, 
      J. Geom. Phys. {\bf 61}  (2011) 2453 $-$ 2469, arXiv:  1012.5807 [hep-th] 
 
 \bibitem{kw18} S. Kondo and T. Watari, {\it String-theory realization of modular forms for elliptic curves with complex multiplication},
    Commun. Math. Phys. {\bf 367} (2019) 89, arXiv: 1801.07464 [hep-th]
 \bibitem{kw19} S. Kondo and T. Watari, {\it Modular parametrization as Polyakov path integral - cases with CM elliptic 
    curves as target spaces}, arXiv: 1912.13294 [hep-th]
    
 \bibitem{k20etal} S. Kachru, R. Nally and W. Yang, {\it Supersymmetric flux compactifications and CY modularity},
      arXiv: 2001.06022 [hep-th]

 \bibitem{d60} B. Dwork, {\it On the rationality of the zeta function of an algebraic variety}, 
 Amer. J. Math. {bf 82} (1960) 631
  \bibitem{w49}  A. Weil, {\it Number of solutions of equations in finite fields}, 
       Bull. Amer. Math. Soc. {\bf 55} (1949) 497 $-$ 508 
 \bibitem{g64} A. Grothendieck, {\it Formule de Lefschetz et rationalite de fonction de $L$}, 
    Seminaire Bourbaki {\bf 279} (1964/65) 1 $-$ 15
 
  \bibitem{rs01} R. Schimmrigk, {\it Arithmetic of Calabi-Yau varieties and rational conformal field theory}, 
      J. Geom. Phys. {\bf 44} (2003) 555 $-$ 569, arXiv: hep-th/0111226 
      
 \bibitem{drs99} K. Dasgupta, G. Rajesh and S. Sethi, {\it M theory, orientifolds and G-flux}, JHEP {\bf 08} (1999) 023, 
        arXiv: hep-th/9908088
  \bibitem{gkp02} S.B. Giddings, S. Kachru and J. Polchinski,  {\it Hierarchies from fluxed in string compactifications},
 Phys. Rev. {\bf D66} (2002) 106006, arXiv: hep-th/0105097  
 \bibitem{grana05} M. Grana, {\it Flux compactifications in string theory: a comphrensive review}, 
    Phys. Rept. {\bf 423} (2006) 91 $-$ 158,  arXiv: hep-th/0509003 
 \bibitem{c07} J.P. Conlon, {\it Moduli stabilization and applications in {\rm IIB} string theory}, 
   Fortschr. Phys. {\bf 55} (2007) 287 $-$ 422,
     arXiv: hep-th/0611039 
 \bibitem{p08}  E. Pajer, {\it Phenomenological aspects of type $\rmII\rmB$ flux compactification},
        Fortschr. Phys. {\bf 57} (2009) 193 $-$ 219

 \bibitem{d04etal}  O. DeWolfe, A. Giryavets, S. Kachru and W. Taylor, {\it Enumerating flux vacua with enhanced 
     symmetries}, JHEP {\bf 02} (2005) 037, arXiv: hep-th/0411061 
     
 \bibitem{cls90} P. Candelas, M. Lynker and R. Schimmrigk, {\it Calabi-Yau manifolds in weighted $\mathP_4$},
     Nucl. Phys. {\bf B341} (1990) 383 $-$ 402
 \bibitem{ks92} A. Klemm and R. Schimmrigk, {\it Landau-Ginzburg string vacua}, Nucl. Phys. {\bf B411} (1994) 559, 
     arXiv: hep-th/9204060 
 \bibitem{krsk92} M. Kreuzer and H. Skarke, {\it No mirror symmetry in Landau-Ginzburg spectra!}, 
      Nucl. Phys. {\bf B388} (1992) 113 $-$ 130, arXiv: hep-th/9205004 
 
 \bibitem{d05}  O. DeWolfe, {\it  Enhanced symmetries in multiparameter flux vacua}, JHEP {\bf 10} (2005) 066, 
     arXiv: hep-th/0506245
 \bibitem{g05} A. Giryavets, {\it New attractors and area codes}, JHEP {\bf 03} (2006) 020, arXiv: hep-th/0511215
     
 \bibitem{c97} J.E. Cremona, {\it Algorithms for modular elliptic curves}, CUP 1997
 \bibitem{m05} C. Meyer,  {\it Newforms of weight two for $\G_0(N)$ with rational coefficients}, December 18, 2005
 \bibitem{lmfdb} LMFDB,  The $L-$function and modular form data base,  http://lmfdb.org
 \bibitem{d79} P. Deligne, {\it Valeurs de fonctions $L$ et periodes d'integrales}, Proc. Symp. Pure Math. {\bf 33} (1979) 313 
 \bibitem{hecke-werke} E. Hecke, {\it Mathematische Werke},  1970 
 \bibitem{rs05} R. Schimmrigk, {\it Arithmetic spacetime geometry from string theory},  
      Int. J. Mod. Phys. {\bf A21} (2006) 6323 $-$ 6350,  arXiv: hep-th/0510091
      
 \bibitem{g03etal} A. Giryavets, S. Kachru, P.K. Tripathy and S.P. Trivedi, {\it Flux compactifications on Calabi-Yau threefolds}, 
      JHEP {\bf 04} (2004) 003, arXiv: hep-th/0312104

\bibitem{m98} G. Moore, {\it Arithmetic and attractors}, arXiv:hep-th/9807087
\bibitem{lps03} M. Lynker, V. Periwal and R. Schimmrigk, {\it Complex multiplication symmetry of black hole attractors}, 
     Nucl. Phys. {\bf B667} (2003) 484 $-$ 504, arXiv: hep-th/0303111
\bibitem{lss03} M. Lynker, R. Schimmrigk and S. Stewart, {\it Complex multiplication of exactly solvable Calabi-Yau varieties}, 
   Nucl. Phys. {\bf B700} (2004) 463 $-$ 489, arXiv: hep-th/0312319 
\bibitem{c20etal} P. Candelas, X. de la Ossa, M. Elmi and D. van Straten, {\it A one parameter family of Calabi-Yau manifolds with 
    attractor points of rank two}, arXiv: 1912.06146  [hep-th]
 \bibitem{b86} D. Blasius, {\it On the critical values of Hecke $L$-series}, Ann. Math. {\bf 124} (1986) 23 $-$ 63
 \bibitem{y20} W. Yang, {\it Rank-2 attractor and Deligne's conjecture}, arXiv: 2001.07211 [math.AG]
 \bibitem{b15etal} R. Blumenhagen, A. Font, M. Fuchs, D. Herschmann, E. Plauschinn, Y. Sekiguchi and F. Wolf, 
     {\it A flux-scaling scenario for high-scale moduli stabilization in string theory}, Nucl. Phys. {\bf B897} (2015) 500 $-$ 554,
      arXiv: 1503.07634 [hep-th]
 \bibitem{ls04} M. Lynker and R. Schimmrigk, {\it Geometric Kac-Moody modularity}, J. Geom. Phys. {\bf 56} (2006) 843, 
       arXiv: hep-th/0410189
 
 
 \bibitem{mn04} A. Misra and A. Nanda, {\it Flux vacua statistics for two-parameter Calabi-Yaus}, 
      Fortschr. Phys. {\bf 53} (2005) 246 $-$ 259, arXiv: hep-th/0407252
 \bibitem{cq04} J.P. Conlon and F. Quevedo, {\it On the explicit construction and statistics of Calabi-Yau flux vacua}, 
            JHEP {\bf 10} (2004) 039, arXiv: hep-th/0409215 
 \bibitem{c07etal} M. Cicoli, J.P. Conlon and F. Quevedo, {\it Systematics of string loop corrections in type IIB Calabi-Yau 
   compactifications}, JHEP {\bf 01} (2008) 052,  arXiv: 0708.1873 [hep-th]
\bibitem{c19etal} N. Cribiori, R. Kallosh, A. Linde and C. Roupec, {\it Mass production of {\rm IIA} and {\rm IIB} dS vacua},
    JHEP {\bf 02} (2020) 063,  arXiv: 1912.00027 [hep-th]
 \bibitem{g04etal} A. Giryavets, S. Kachru and P.K. Tripathy, {\it On the taxonomy of flux vacua}, JHEP {\bf 08} (2004) 002, 
        arXiv: hep-th/0404243
\bibitem{s97} A. Sen, {\it Orientifold limit of F-theory vacua}, Phys. Rev. {\bf D55} (1997) 7345, arXiv: hep-th/9702165


\bibitem{lsw98} M. Lynker, R. Schimmrigk and A. Wisskirchen, {\it Landau-Ginzburg vacua of string, M-theory, and F-theory 
 at $c=12$},  Nucl. Phys. {\bf B550} (1999) 123 $-$ 150, arXiv: hep-th/9812195
\bibitem{ellcy4list} M. Lynker, R. Schimmrigk and A. Wisskirchen, {\it Data archive for Calabi-Yau fourfolds}: \\
http://www.th.physik.uni-bonn.de/Supplements/cy.html 
 \bibitem{cs18} A. Cole and G. Shiu,  {\it Topological data analysis for the string landscape}, JHEP {\bf 03} (2019) 022, 
  arXiv: 1812.06960 [hep-th]
 \bibitem{css19} A. Cole, A. Schachner and G. Shiu, {\it Searching the landscape of flux vacua with genetic algorithms},
           JHEP {\bf 11} (2019) 045, arXiv:1907.10072 [hep-th]
     
\bibitem{ddf04} F. Denef, M.R. Douglas and B. Florea, {\it Building a better racetrack}, JHEP {\bf 06} (2004) 034, 
      arXiv:  hep-th/0404257 
\bibitem{b05etal} V. Balasubramanian, P. Berglund, J. Conlon and F. Quevedo, {\it Systematics of moduli stabilisation 
     in CY flux compactifications}, JHEP {\bf 03} (2005) 007, arXiv: hep-th/0502058 
\bibitem{c05etal} J.P. Conlon, F. Quevedo and K. Suruliz, {\it Large-volume compactifications: moduli spectrum and D3/D7 
  soft supersymmetry breaking}, JHEP {\bf 08} (2005) 007, arXiv: hep-th/0505076
\bibitem{ds05}  M. Dine and Z. Sun, {\it R symmetries in the landscape}, JHEP {\bf 01} (2006) 129, arXiv: hep-th/0506246 
\bibitem{b06etal} J.J. Blanco-Pillado, C.P. Burgess, J.M. Cline, C. Escoda, M. Gomez-Reino, R. Kallosh, A. Linde and F. Quevedo,
       {\it Inflating in a better racetrack},  JHEP {\bf 09} (2006) 002, arXiv: hep-th/0603129  
\bibitem{a07etal} S.S. AbdusSalam, J.P. Conlon, F. Quevedo and K. Suruliz, {\it Scanning the landscape of flux compactifications: 
     vacuum structure and soft supersymmetry breaking}, JHEP {\bf 12} (2007) 036, arXiv: 0709.0221 [hep-th]
\bibitem{b09etal} R. Blumenhagen, V. Braun, T.W. Grimm and T. Weigand, {\it GUTs in type IIB orientifold compactifications}, 
        Nucl. Phys. {\bf B815} (2009) 1 $-$ 94, arXiv: 0811.2936 [hep-th]
\bibitem{l12etal} J. Louis, M. Rummel, R. Valandro and A. Westphal, {\it Building an explicit de Sitter}, JHEP {\bf 10} (2012) 163,
     arXiv: 1208.3208 [hep-th]
\bibitem{d19etal} M. Demirtas, M. Kim, L. McAllister and J. Moritz, {\it Vacua with small flux superpotential}, 
       arXiv: 1912.10047 [hep-th]
     
\bibitem{et05} T. Eguchi and Y. Tachikawa, {\it Distribution of flux vacua around singular points in Calabi-Yau moduli space}, 
      JHEP {\bf 01} (2006) 100,  arXiv: hep-th/0510006 
\bibitem{pw14} N.M. Paquette and T. Wrase, {\it Comments on $M_{24}$ representations and CY$_3$ geometries},
   JHEP {\bf 11} (2014) 155, arXiv: 1409.1540 [hep-th] 
   
\bibitem{mnp13} J.P. Murre, J. Nagal and C.A.M. Peters, {\it Lectures on the theory of pure motives},  University Lecture 
    Series, AMS 2013
   \bibitem{a04} Y. Andr\'e, {\it Une introduction aux motifs
  (Motif purs, motifs mixtes, p\'eriodes}), Soci\'et\'e Math\'ematique de France,  2004
  
 \end{thebibliography}
 \end{document}